\documentclass{optica-article}
\journal{oe}

\articletype{Research Article}

\usepackage{lineno}

\begin{document}
\title{Fluorescent property of carbon dots extracted from cigarette smoke 
and the application in bio-imaging}

\author{Yuzhao Li,\authormark{1} Han Bai,\authormark{1,2} 
Jin Zhang,\authormark{1,3,5} Ju Tang,\authormark{4,6} 
Yingfu Li,\authormark{1} Weizuo Zhang,\authormark{1} 
Zhexian Zhao,\authormark{1} Yiming Xiao,\authormark{1,7} 
and Yanfei L\"{u}\authormark{1,*}}

\address{
\authormark{1}School of Physics and Astronomy, Yunnan University, Kunming 650091, China\\
\authormark{2}Department of Radiation Oncology, Yunnan Tumor Hospital, Kunming 650106, China\\
\authormark{3}Yunnan Carbon Based Technology Co. LTD, Kunming 650028, China\\
\authormark{4}Department of Physics, School of Electrical and Information
Technology, Yunnan Minzu University, Kunming 650504, China}
\email{\authormark{5}zhangjin@ynu.edu.cn} 
\email{\authormark{6}tangju@ymu.edu.cn} 
\email{\authormark{7}yiming.xiao@ynu.edu.cn} 
\email{\authormark{*}optik@sina.com} 


\begin{abstract}
Cigarette smoke is one of the six major pollution sources in the room air. 
It contains large number of particles with size less than 10 nm. There 
exist carbon dots (CDs) in cigarette smoke which have strong fluorescence 
and with good bio-compatibility and low toxicity. CDs in cigarette smoke 
can be applied in bio-imaging which has great potential applications in 
the integration of cancer diagnosis and treatment. In this paper, CDs were 
extracted from cigarette smoke. Then, sodium borohydride was added to CDs 
aqueous solution for reduction and the reduced CDs (R-CDs) were used for
biological cell imaging. The results indicate that the CDs with the
particle size $<$10 nm in cigarette smoke are self-assembled by the
polymerizated polycyclic aromatic hydrocarbons (PAHs) and ammonium nitrite
which are disk nano-structure composed of $sp^2$/$sp^3$ carbon and
oxygen/nitrogen groups or polymers. Sodium borohydride can reduce the
carbonyl group on the surface of CDs to hydroxyl group and increase the ratio 
of the Na 1s ratio of the CDs from 1.86 to 7.42. The CDs can emit blue 
fluorescence under ultraviolet irradiation. After reduction, the R-CDS have 
the intensity of fluorescence 7.2 times than before and the fluorescence 
quantum yield increase from 6.13\% to 8.86\%. The photoluminescence (PL) 
wavelength of R-CDS have red-shift of 7 nm which was due to the increasing 
of Na element ratio. The onion epidermal cells labeled with R-CDs show that 
the CDs could pass through the cell wall into the cell and reach the nucleus. 
The cell wall and the nucleus could be clearly visualized. CDs also shows 
low toxicity to human bronchial epithelial cells (BEAS-2B) with good biological 
activity. The obtained results indicate that the CDs and R-CDs have good 
fluorescent property which could be used as bio-imaging agent.
\end{abstract}

\section{Introduction}
Carbon dots (CDs) are new types of zero-dimensional (0D) carbon-based 
nanomaterials with excellent properties such as strong fluorescence, good 
water solubility, low toxicity, good photostability, chemical stability, and
bio-compatibility \cite{Lu21,Lin21,Kalanidhi21,Bai22}. Therefore, it has great
potential applications in light emitting devices, green lighting, biological 
imaging, cell labeling, photocatalysis, and many other fields \cite{Pan21,Zhang21}. 
CDs can be prepared from abundant raw materials which can be classified as 
biomass and non-biomass. Biomass raw materials include plants, animals and 
their derivatives or wastes \cite{Qureshi21}. Non-biomass raw materials are 
chemical reagents, fossil fuels, and minerals such as coal and graphite \cite{Zhang21J}.
However, biomass carbon sources can be easily gained from nature with which
are diverse and renewable. Therefore, the investigations of CDs made from 
biomass has attracted lots of attentions.

It is worth noting that the ultra-fine particles with size $<$10 nm
generated by the combustion of biomass and fossil energy are very common
such as cigarette smoke, soot from coal-fired power generation, smoke
from forest fires, volcanic ash, and etc. These nanoscale particles 
cause serious pollution to the environment of human life and are harmful 
to human health. However, some relevant studies only focus on the impact 
of particles with size $>$20 nm on human
health \cite{Keith82,Morie77,Carter75,Morawska97,Czoli15,Benner89,Ning06,
Miller01,Hinds78,Nazaroff93} whereas the effect of CDs with size $<$10 nm
have not received enough attentions \cite{McCusker83,Morawska99,Li14}.

Tobacco contains more than 7000 kinds of chemical substances and about
400 of them can cause toxicity to human body with more than 50 kinds of
carcinogens including arsenic, cadmium, formaldehyde, and benzo pyrene \cite{Baker90}.
Cigarette smoke is the most common tobacco product. It's one of the main
sources of human exposure to fine particles in the air. 92\% of cigarette
smoke is gas which mainly contains carbon monoxide, carbon dioxide, nitrogen
oxides, volatile low molecular alkanes, and alkenes, etc. Besides, the other
8\% is particulate matters (PM) \cite{Zheng18,Bernstein04,Zhang21ER} which 
is smoke dust with particle size of 0.1-2 $\mu$m and an average diameter 
of 0.2 $\mu$m. The smoke dust becomes tar after condensation. Each cigarette 
produces 6.3$\times10^9$ particles per second which are about 20-35 mg tar. 
It has been confirmed that they contain more than 4000 carcinogens such as 
nicotine, polycyclic aromatic hydroxyl, benzo pyrene, and 
$\beta$-naphthylamine \cite{Eatough89}. Carbonyl compounds are main components 
of cigarette smoke. The carbonyl particles in cigarette smoke account for 
11-19\% of the total particulate mass and some of them are toxic and 
may be carcinogenic or mutagenic to human being \cite{Pang11}.

When the cigarettes are burning, the surface temperature of the burning
area is about 300 $^{\circ}$C and the central combustion temperature is
about 900 $^{\circ}$C \cite{Norman99,Zheng06}. The organic matters in tobacco
will produce a large number of fluorescent CDs with the particle size of
7 nm-10 $\mu$m \cite{Nicolaou21}. The number of PM with particle size
$\leq$ 0.1 $\mu$m is roughly the same as that of 0.1-2.0 $\mu$m \cite{Ma22}.
The PM with particle size of 0.1-2.0 $\mu$m contribute to the most of the total
PM mass. A large number of CDs with size $\leq$10 nm are often overlooked 
because they are difficult to detect with small mass and size.

In order to explore the structure of particles with size $<$10 nm in
cigarette smoke and the role of surface carbonyl and hydroxyl groups played, 
CDs were extracted from the static burning smoke of 60 cigarettes made in China.
The structure, chemical composition, surface groups, and fluorescence characteristics
of CDs were characterized by TEM, XPS, XRD, PL, absorption spectrum, and particle size
distribution. The carbonyl group in CDs was reduced by NaBH$_4$ and converted to
hydroxyl group to passivate the CDs and enhance the fluorescence. Meanwhile, the R-CDs 
are injected into onion epidermal cells to study the beneficial effect of R-CDs on 
the imaging quality of plant cells. The toxicity test is performed to evaluate 
the toxicity of CDs on BEAS-2B.

\section{Experiment}
\label{sec:Experiment}

\subsection{Materials and experimental instruments}
The cigarettes (Yunyan brand) were purchased from local supermarket. Ethanol and sodium
borohydride are pure analytical reagents. The deionized water used in the experiment was
produced by Master-S15 deionized water production machine (Shanghai Hetai Instrument
Co., LTD). Cigarette smoke collection equipment includes SHZ-III recirculating water 
vacuum pump (Shanghai Yarong Biochemical Instrument Factory), TG16G table top centrifuge 
(Yancheng kaite Experimental Instrument Co., LTD.), and KUDOS ultrasonic cleaning 
equipment (Shenzhen Xinbao Instrument Co., LTD.). The fluorescence spectrum
of CDs was measured by a fluorescence spectrometer F9818012 (Shanghai lenguang 
technology Co., LTD). The absorption spectra were measured by a UV-Vis spectrophotometer 
(Specord200, Germany). The morphology and micro-structures of the CDs were characterized 
by using the transmission electron microscopy (TEM, JEM 2100). The X-ray photoelectron 
spectroscopy (XPS) of CDs were measured by using PHI5000 Versa Probe II photoelectron 
spectrometer with Al $K_{\alpha}$. The infrared absorption spectrum was measured by a 
Fourier transform infrared spectrometer (Nicoletis10, USA).

\subsection{Synthesis of CDs with the cigarette}
Firstly, 400 ml of deionized water and absolute ethanol were injected into 
two elution bottles, respectively. Absolute ethanol 
was used as the solvent to dissolve tar in flue gas, and deionized water was
used as the solution to extract CDs in flue gas. Then, 60 cigarettes (Yunyan brand)
burns in turn and place them at one of the end of the vent. Cigarette smoke
is pumped into the bottle through a vacuum pump. CDs are extracted
with deionized water by using the good water solubility of CDs. The CDs solution
was a light yellow liquid. After centrifuged and filtered by a filter
membrane with a pore size of 220 nm, the CDs solution was stored in a
refrigerator for later use.

In order to enhance the intensity of fluorescence of the CDs, the CDs was 
passivated. 10 ml of the collected CDs solution with a concentration
of 0.2 mg/mL and 60 mg sodium borohydride (NaBH$_4$) are added to the CDs 
aqueous solution and it was stirred at room temperature for full reaction. 
After reaction, the CDs solution was filtered by a filter membrane with a 
pore size of 220 nm.

\subsection{Cell imaging study and the toxicity of CDs}
The inner membrane of onion epidermis was torn off with tweezers from the
fresh onions. CDs was dropped into the epidermis for 5 mins with natural
absorption. Then, the structure of onion epidermis cells was observed under
ultraviolet, violet, and green light with a fluorescence microscope.
In order to evaluate the toxicity of CDs, BEAS-2B cultured to 
logarithmic growth stage are implanted into 96-well culture plate with 
2000 cells per well. 24 hours later, the cells are stuck to the wall. Then, 
suck out them from the medium and added 200 $\mu$L/well CDs solution with 
different concentrations (0.08-1.38 mg/mL) and the solvent is
serum-free cell medium. After culturing for 24 hours, CCK-8 assay is used 
to measure the cell activity.

\begin{figure}[b]
\begin{center}
\includegraphics[width=0.8\linewidth]{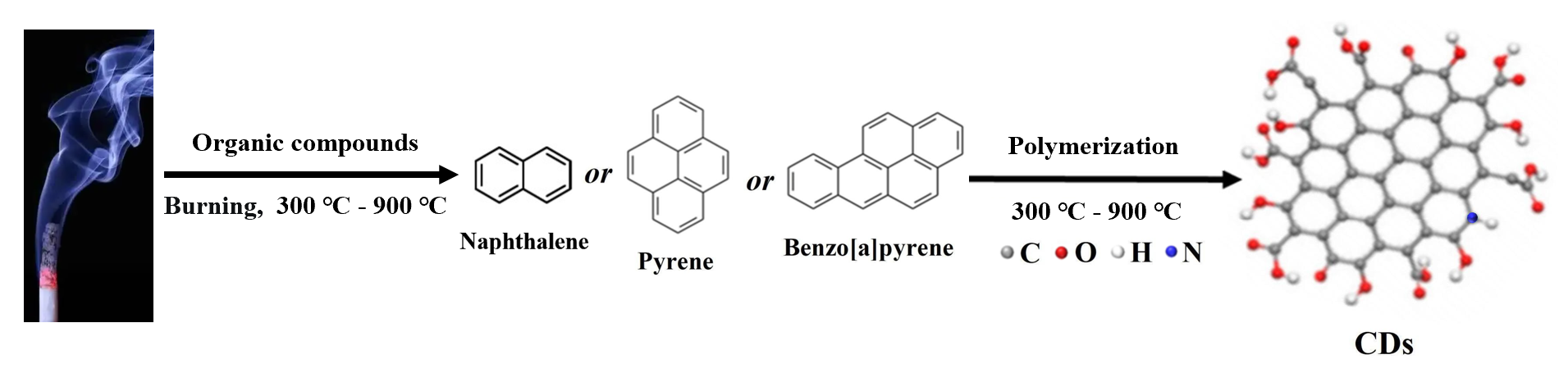}
\end{center}
\caption{The possible formation mechanism and structural characteristics 
of CDs in smoke.}\label{fig1}
\end{figure}

\section{Results and Discussions}
The particulate matter (PM) produced by cigarette 
combustion usually contains polycyclic aromatic hydrocarbons (PAHs), 
organic nitrates, ethylpyridine, and other organic compounds. PAHs are 
mainly produced from incomplete combustion of organic matter. They widely 
exist in environmental pollutants worldwide and are known for their 
carcinogenic, mutagenic, and teratogenic properties. Among more than 
500 carcinogenic compounds, PAHs and its derivatives account for more 
than 200. The ratio of PAHs in PM is about 57 \% while naphthalene (C$_{10}$H${_8}$) is 
the simplest and most abundant PAHs in smoke \cite{Zhang21ER}. During 
the burning of cigarettes, organic matter in tobacco is firstly pyrolysed 
to PAHs and then polymerized during the burning process to form a large 
number of CDs with size $<$10 nm. Fig. \ref{fig1} shows the possible 
formation mechanism and structural characteristics of CDs in cigarette smoke. 
PAHs and ammonium nitrite contained 
in the tar of cigarette smoke are polymerized and self-assembled into 
disk nano-structure composed of $sp^2$/$sp^3$ carbon crystal nuclei and 
oxygen/nitrogen group or polymer. The structure diagram of CDs is shown in 
the right side of Fig. \ref{fig1}.

\begin{figure}[t]
\begin{center}
\includegraphics[width=0.75\linewidth]{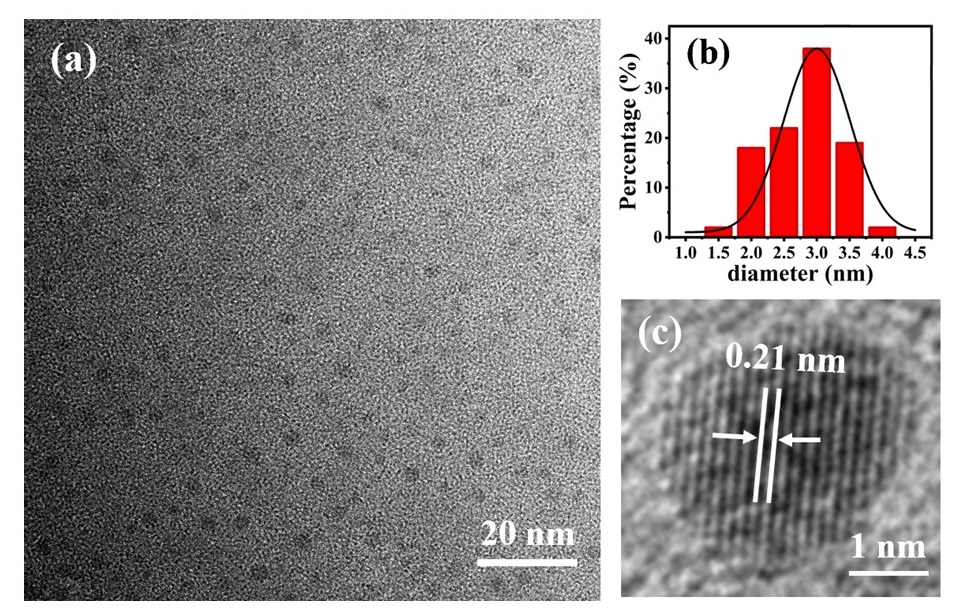}
\end{center}
\caption{The TEM image (a) and particle size distribution diagram (b) 
of CDs in cigarette smoke. (c) the HRTEM image of the CDs' lattice.}
\label{fig2}
\end{figure}

\begin{figure}[b]
\begin{center}
\includegraphics[width=0.6\linewidth]{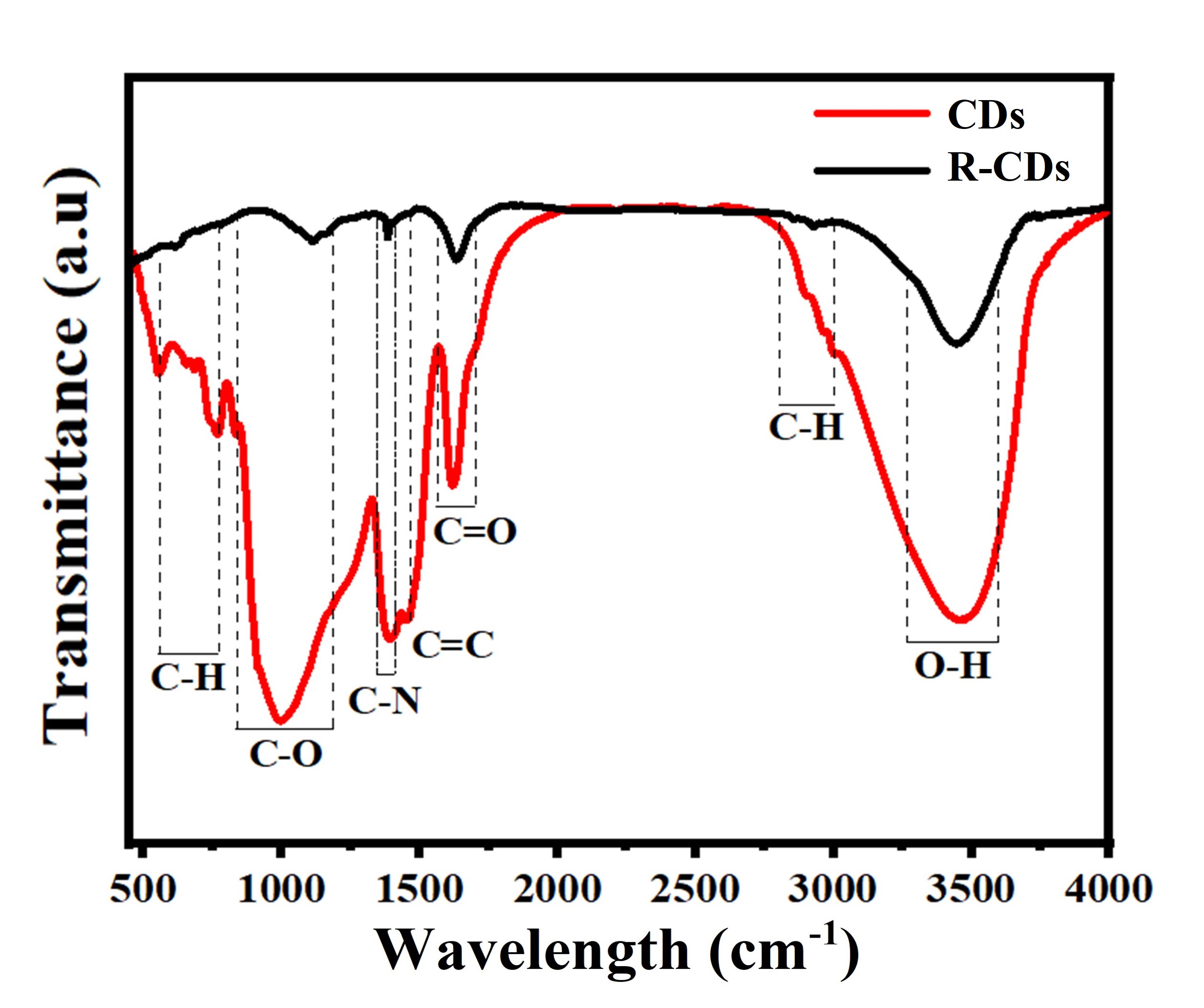}
\end{center}
\caption{Infrared absorption spectra of CDs
and R-CDs (after modifications with NaBH$_4$). The region of 
the vibrations for different bonds are indicated.}
\label{fig3}
\end{figure}

In Fig. \ref{fig2}, we show the TEM image, particle size distribution of CDs, and
HRTEM image of CDs' lattice. As can be seen from Fig. \ref{fig2} (a), the
morphology of CDs is quasi-circular and with good dispersion. Fig. \ref{fig2}
(b) shows that the diameter of CDs is between 1-4 nm. The diameter particle
distribution is close to normal distribution with an average diameter of
3 nm. Fig. \ref{fig2} (c) shows that the core region of a single CD
is crystalline structure with clear lattice fringes. The crystal plane 
spacing is about 0.21 nm, which is corresponding to the (100) crystal plane of
graphene \cite{Zhang21SA,Hasrudin21}.

\begin{figure}[t]
\begin{center}
\includegraphics[width=0.8\linewidth]{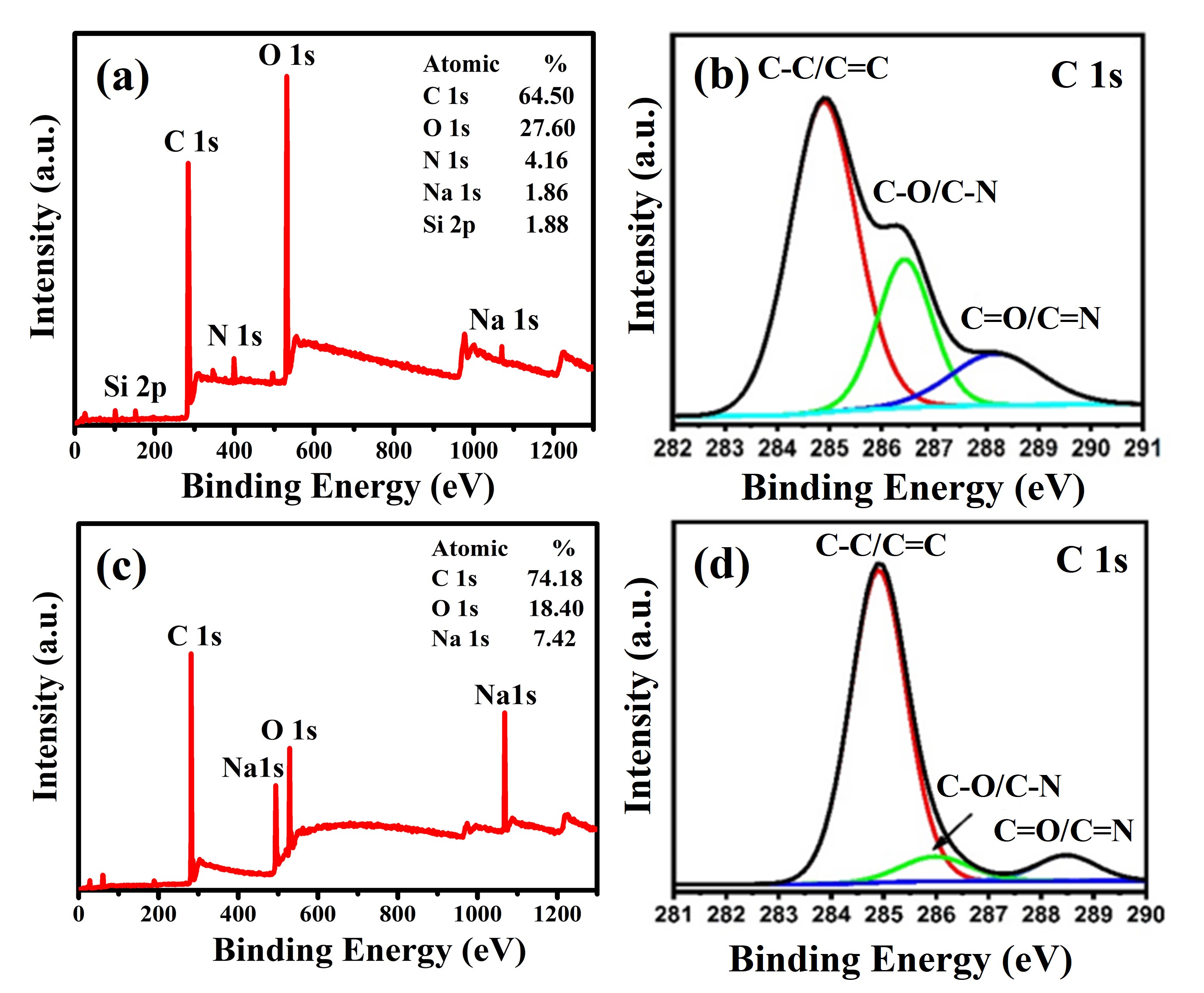}
\end{center}
\caption{The full XPS spectrum (a) and C1s peak fitting diagram (b) of CDs.
The full XPS spectrum (c) and C1s peak fitting diagram (d) of R-CDs.}
\label{fig4}
\end{figure}

In order to study the changes of carbon skeleton structure and surface groups
after adding NaBH$_4$ modifier to CDs solution, the samples were measured by
Fourier infrared absorption spectroscopy (FTIR) and X-ray photoelectron spectroscopy 
(XPS). Fig. \ref{fig3} shows the infrared absorption spectra of CDs solution
and R-CDs solution after reduction by NaBH$_4$. As we can seen in Fig. \ref{fig3}, 
the absorption peaks of these two curves are approximately the same. The peaks at 
3445 cm$^{-1}$ and 2925 cm$^{-1}$ belong to the stretching vibration peak of O-H 
and C-H bonds, respectively. Peaks at 1586 cm$^-1$, 1384 cm$^{-1}$, 1358 cm$^{-1}$, 
and 966 cm$^{-1}$ were generated by stretching vibration of C=O, C=C, C-N, and C-O 
bonds, respectively \cite{Li21,Mu20}. The peaks at 737 cm$^{-1}$ and 527 cm$^{-1}$ 
are caused by the C-H stretching vibration \cite{Hu22}. After NaBH$_4$ was added to 
the CDs solution, CDs is reduced to R-CDS. As can be seen from Fig. \ref{fig3}, 
the C=C bond in the carbon core region increase significantly with the enhancement 
of the stretching vibration of O-H bond. Meanwhile, C-O/C=O also increase significantly 
which indicates that the number of hydroxyl groups on the surface of R-CDs is 
significantly more than that of carbonyl groups. This is beneficial to the improvement 
of the yield and fluorescence lifetime of CDs \cite{Tang19,Tang21}.

Fig. \ref{fig4} (a) shows the full spectrum of XPS of CDs extracted from PM
dispersed in deionized water filtered from cigarette smoke. The XPS analysis 
of CDs shows that the CDs without the modification of NaBH$_4$ mainly consists of 
three elements, C, O, and N, with the atomic percentages of 64.5\%, 27.6\%, and 4.16\%, 
respectively. It also contains small amounts of Na and Si which are due to the metal 
in the cigarette and the Si element in the quartz glass used as the test substrate.
Fig. \ref{fig4} (b) shows the C1s peak fitting diagram of CDs. The C1s peak can be divided
into three peaks at 284.9 eV, 286.4 eV, and 288.3 eV which correspond
to C-C/C=C, C-O/C-N and C=O/C=N bonds, respectively \cite{Stepanova21}.
Fig. \ref{fig4} (c) shows the full spectrum of XPS after the reduction of CDs extracted
from cigarette smoke by NaBH$_4$. The R-CDs after reduction are mainly composed of
C and O, and their atomic percentages are C: 74.18\% and O: 18.4\%, respectively.
Other trace elements are Na and Cl in flue gas and Si in the quartz glass. 
After CDs was reduced by NaBH$_4$, C and Na element increases and O element decreases. 
The content of nitrogen element in CDs becomes zero. Fig. \ref{fig4} (d) is the 
peak fitting diagram of C1s of R-CDs. There are three peaks located at 284.9eV, 285.9 eV, 
and 288.5 eV which belong to C-C, C-O and C=O bonds, respectively \cite{Wang13}. The 
results indicate that NaBH$_4$ can reduce the carbonyl group on the surface of the 
CDs to hydroxyl group and increase the Na 1s ratio of the CDs group from 1.86 to 
7.42. Thus, the CDs can be passivated and it can reduce the toxicity of CDs. The 
XPS results are in line with the FTIR results. 

\begin{figure}[t]
\begin{center}
\includegraphics[width=0.85\linewidth]{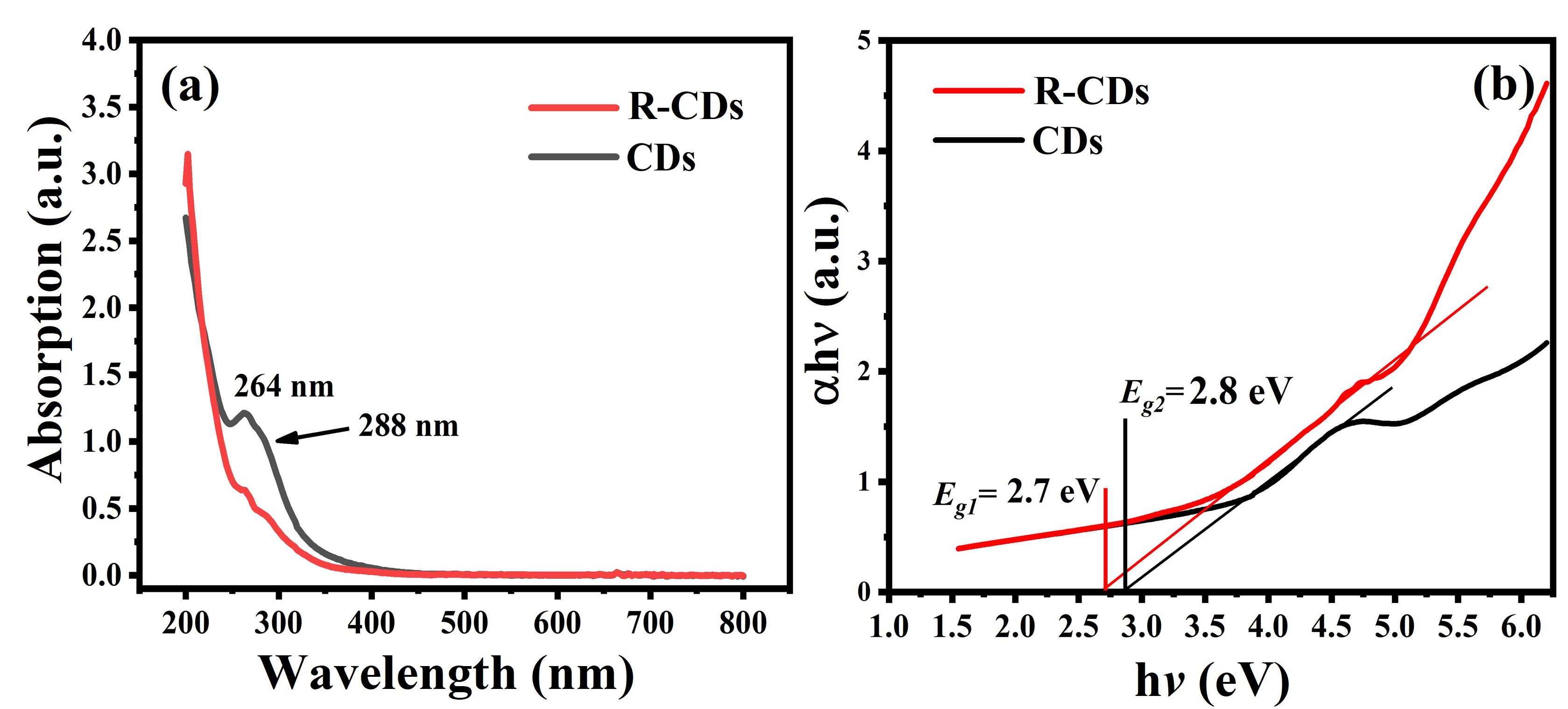}
\end{center}
\caption{UV-vis absorption spectra (a) and the optical band-gap (b)
of CDs and R-CDs.}
\label{fig5}
\end{figure}

Fig. \ref{fig5} (a) shows the absorption spectra of CDs and R-CDs (after reduction).
The CDs before reduction has an obvious absorption peak at 264 nm and a weak shoulder
peak at 288 nm. The absorption peak at 264 nm is attributed to the $\pi-\pi^*$ transitions
generated by C=C in aromatic hydrocarbons. The shoulder peak at 288 nm is the absorption
induced by the $n-\pi^*$ transitions generated by C=O bond \cite{Tiwari20}. For 
the R-CDs, the intensity of the absorption peak is obviously weakened while the peak position
remains unchanged. Fig. \ref{fig5} (b) shows the value of optical band-gap
of CDs which is calculated with the relationship between absorption coefficient
and band-gap through $(\alpha hv)^{1/2}=hv-E_g$, where $\alpha$ is absorption coefficient,
$hv$ is photon energy, $E_g$ is band gap. The optical band-gaps of CDs and R-CDs
are estimated to be 2.8 eV and 2.7 eV, respectively. The linear region of the
indirect band-gap semiconductor absorption spectrum has a shoulder structure which means
that the calculated band-gap parameters are for indirect band-gaps.

\begin{figure}[t]
\begin{center}
\includegraphics[width=0.7\linewidth]{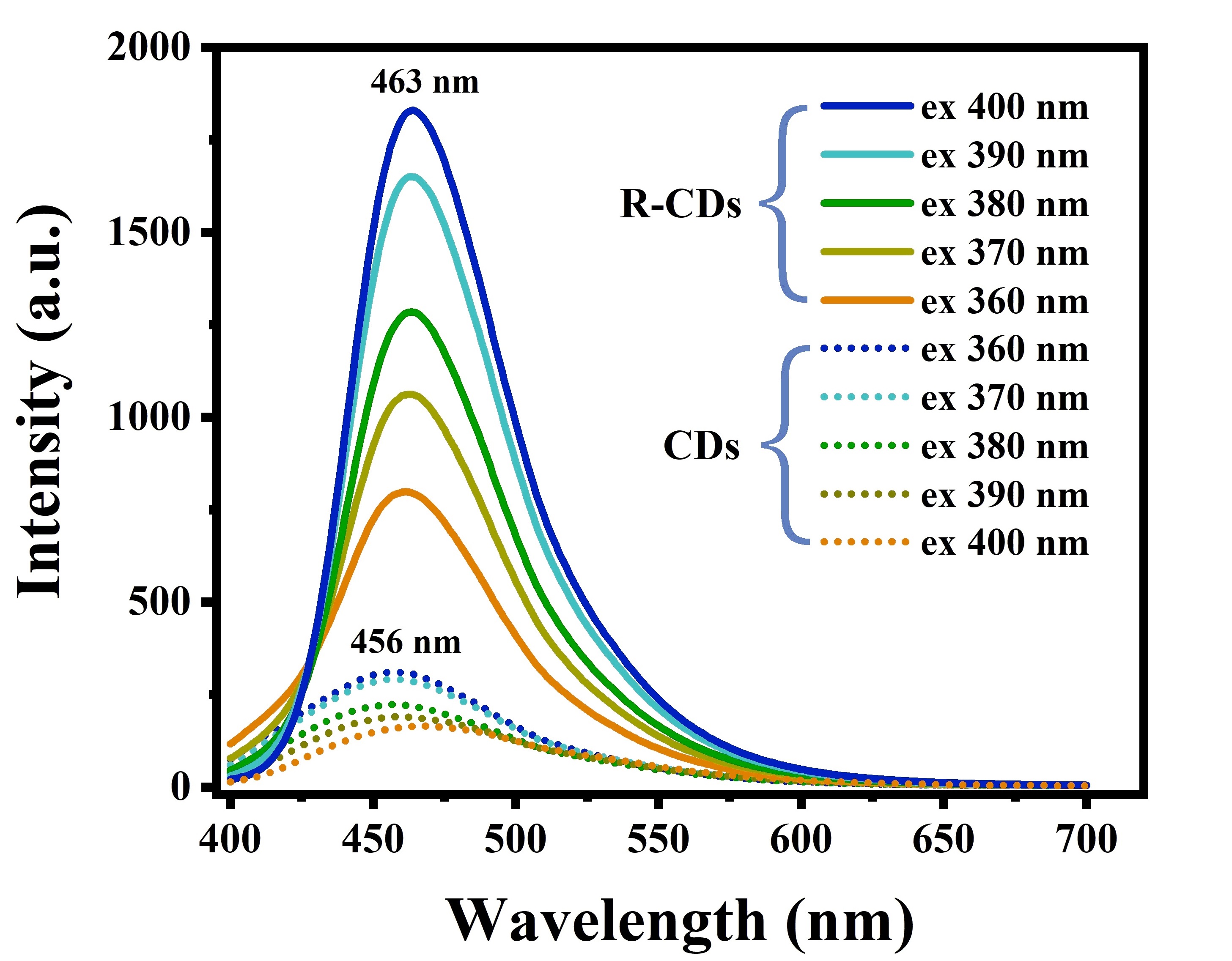}
\end{center}
\caption{Fluorescence spectra of CDs and R-CDs in cigarette smoke.}
\label{fig6}
\end{figure}

In Fig. \ref{fig6},  the solid curves show the fluorescence emission spectrum of CDs. 
It can be seen that the emission range is between 400-700 nm and the peak position 
is at 456 nm. The peak of emission spectrum does not change with the excitation wavelength. 
The dashed curves show the
fluorescence emission spectra of R-CDs reduced by sodium borohydride in order to
enhance the fluorescence intensity of CDs. The fluorescence 
intensity can be calculated through the integral area of the fluorescence curve.
We can see that the intensity of the
fluorescence emission with an excitation wavelength 360 nm is significantly enhanced 
with 7.2 times after reduction.
This is due to the reduction of carbonyl group to hydroxyl group by NaBH$_4$,
which reduces the non-radiative transition caused by carbonyl group and
thus increases the fluorescence intensity. By the way, the optimal excitation
wavelength and fluorescence emission peak are red-shifted which is due 
to the increasing of Na element ratio.  This is analogous to 
a doped semiconductor, where the impurity levels $E_D$ and $E_A$ in the forbidden 
band correspond to the donor and acceptor energy levels, respectively. Here, $E_D$ is 
equivalent to surface states, which are mainly induced by the surface groups and 
donor impurities of CDs or R-CDs. $E_A$ is the acceptor energy level formed by the 
metal ion Na$^+$ in R-CDs after reduction of NaBH$_4$. When the percentage of Na 
elements increase from 1.86 \% in CDs to 7.42 \% in R-CDs, the width of the impurity 
level formed by Na$^+$ increases and the energy gap between $E_D$ and $E_A$ decreases. 
This could result in a 7 nm red-shift of the fluorescence wavelength of R-CDs compared
with CDs \cite{Sun21}.

\begin{figure}[t]
\begin{center}
\includegraphics[width=0.9\linewidth]{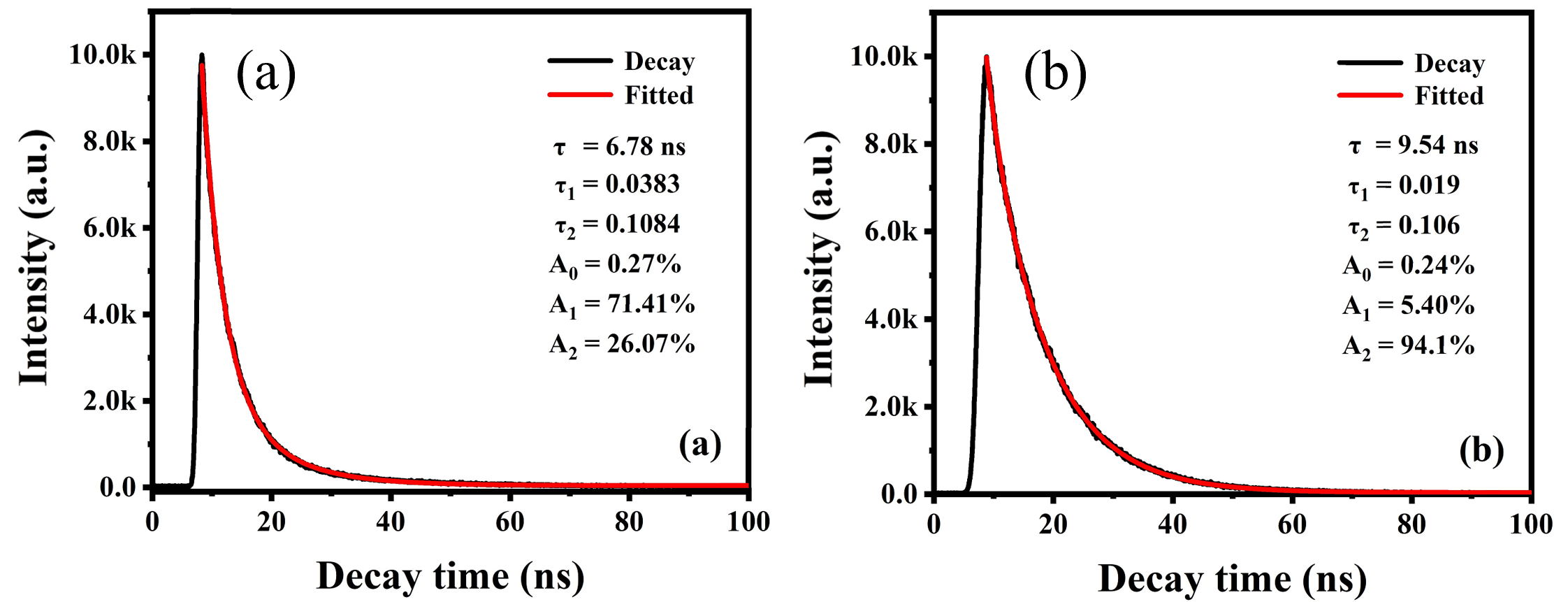}
\end{center}
\caption{The fitted curves of fluorescence lifetime of (a) CDs and (b) R-CDs.}
\label{fig7}
\end{figure}

A combined fluorescence lifetime and steady-state fluorescence spectrometer
(Edinburgh Instruments, UK) equipped with an integrating sphere was used to measure
the fluorescence quantum yield and fluorescence decay curve at room temperature. 
Fig. \ref{fig7} shows the fluorescence lifetime decay curves of 
CDs and R-CDs. The fluorescence decay can be fitted by a double-exponential decay 
model through
\begin{equation}\label{1}
R(t)=A_0+A_1\exp{(-t/t_1)}+A_2\exp{(-t/t_2)},
\end{equation}
where $A_0$ is the background 
PL intensity and $A_1$ and $A_2$ are the fractional contributions to PL emission from 
two transition channels with corresponding decay time or lifetime. Through fitting, the 
fluorescence lifetimes of CDs and R-CDs are 6.78 ns and 9.54 ns, respectively. The Channel-1 
with CDs contributes 71.41 \% of the PL emission, while the Channel-2 contributes 26.07 \%. 
The Channel-1 with R-CDs contributes 0.54 \% of the PL emission, while the Channel-2 contributes 
94.1 \%. Specifically, Channel-1 relates to the transition between the intrinsic 
carbon base states in CDs, while Channel-2 corresponds to the transition with the surface 
states in CDs via non-radiation and radiative electronic relaxations \cite{Tang21}. Such 
a result indicates that the radiative electronic transition can be acquired in nanoseconds 
in CDs. By the way, the width of the impurity level formed by Na$^+$ would also attribute 
to the non-radiative process and the fluorescence process of R-CDs is more complicated 
than that of CDs \cite{Ghorai22}. Thus, the fluorescence lifetime of R-CDs is longer 
than that of CDs. Moreover, the fluorescence quantum yield of CDs is calculated through
\begin{equation}\label{2}
Q=Q_{S}\cdot \frac{I_{S}}{I}\cdot \frac{A}{A_{S}}\cdot \frac{\eta}{\eta_{S}},
\end{equation}
where $Q_{S}$ is the quantum yield of the fluorescence for a standard sample for
reference. $I$ and $I_{S}$ are the integrated emission intensities of the CDs sample and the standard
sample, respectively. $A$ and $A_{S}$ are respectively the absorbance of the prepared
sample and standard sample at the same excitation wavelength. $\eta$ and $\eta_{S}$ are
respectively the refractivity of the prepared sample and standard sample.
In this study, the fluorescence quantum yield of CDs is calculated relatively
to the standard fluorescent substance quinine sulfate. The instrument used in this
experiment has been calibrated with quinine sulfate standard solution during system
setup. Thus, the fluorescence quantum yield of CDs can be measured by inserting
the sample into the integrating sphere. The quantum yield of CDs and R-CDS are 6.13\% and 8.86\%,
respectively.

\begin{figure}[b]
\begin{center}
\includegraphics[width=0.9\linewidth]{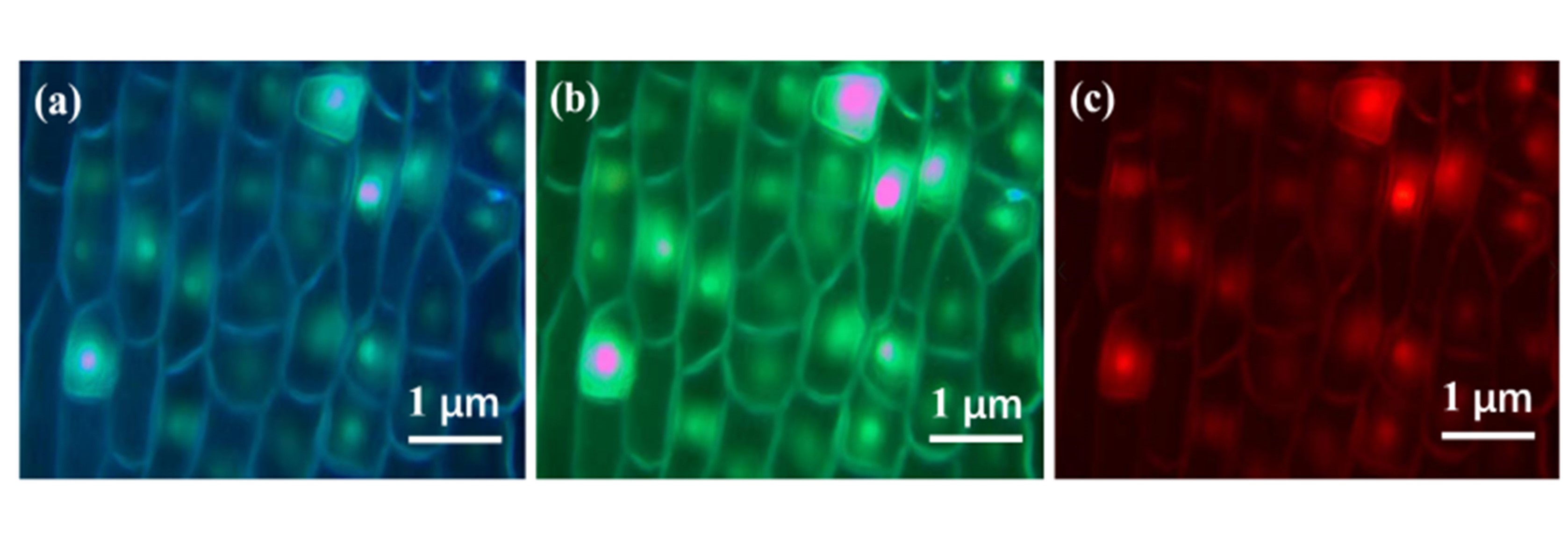}
\end{center}
\caption{Fluorescence microscopic images of onion epidermal cells
under different excitation light irradiation were labeled. (a) Ultraviolet
light exposure, (b) violet light exposure, and (c) green light exposure, respectively.}
\label{fig8}
\end{figure}

\begin{figure}[t]
\begin{center}
\includegraphics[width=0.8\linewidth]{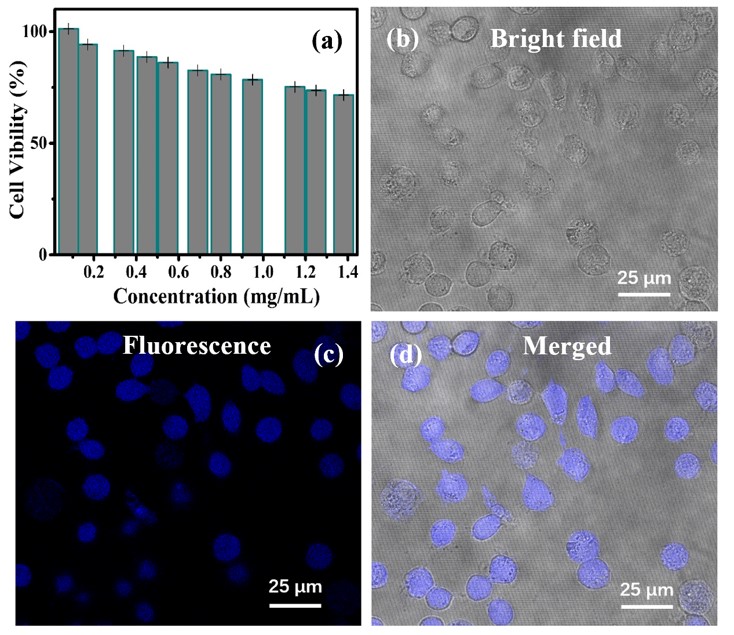}
\end{center}
\caption{Cytotoxicity of CDs on on human bronchial epithelial cells (BEAS-2B) and 
its microscopic image. (a) The cell activity was measured by CCK-8 assay after 24 hours 
of cell culture with gradually increasing CDs concentration from 0.08 mg/mL to 1.38 mg/mL.
The ordinate is the percentage of cell activity in CDs solution compared with that of the 
same amount of normal cells. (b) The bright field image were observed under Leica TCS SP8 
STED laser confocal microscope after adding 0.34 mg/mL CDs solution into the culture medium 
for 4 hours. (c) The fluorescence imaging of BEAS-2B obtained by laser excitation of the 
same sample with wavelength of 405nm. (d) The combination of (b) and (c).}
\label{fig9}
\end{figure}

In Fig. \ref{fig8}, we show micrograph of onion epidermal cells with blue fluorescence under
(a) ultraviolet light (330-400 nm), (b) green fluorescence under violet light (395-415 nm),
and (c) red fluorescence under green light (460-550 nm). It's obvious to see that
the R-CDs can pass through the cell wall into the cell and reach the nucleus.
The cell wall and nucleus are clearly visualized which indicates that the CDs can be
used as a bio-imaging agent. We can also see that there exists pink light
emission in certain cells. However, CDs contains large number of functional groups, 
such as -OH and -COOH. Usually, CDs are sensitivity to the pH of their micro-environment. 
When confronted with acid or alkali environment, the fluorescence of CDs will be greatly 
affected, making its fluorescence red shift or blue shift. Onion epidermis cell structure 
includes cell membrane, cytoplasm, nucleus, cell wall and vacuole. There are a variety 
of substances dissolved in the cell fluid within the vacuoles, and the pH values within 
each cell could vary. Within a specific pH change, the functional group can be altered 
and results in a change in fluorescence intensity and wavelength, e.g., the pink 
light emission \cite{Li22}. Therefore, CDs can also be used as fluorescent 
probe to detect cell variation.

In Fig. \ref{fig9}, we can see that the activity of BEAS-2B depends sensitively 
on the CDs concentration. The cell activity decreases with increasing CDs concentration. 
When the concentration of CDs is less than or equal to 0.08 mg/mL, the cell activity is 
almost not affected. However, when the concentration of CDs was great than or equal to 
1.38 mg/mL, the cell activity is below 70 \% as shown in Fig. \ref{fig9}(a). Moreover, BEAS-2B 
are grown to a sufficient amount great than 1000 in a cell culture flask and transplant 
them into a laser confocal dish for culture. After the cells stuck to the wall, the medium 
is sucked out and adding the CDs solution with a concentration of 0.34 mg/mL with the complete 
medium solvent. After 4 hours, the bright field image of BEAS-2B measured by a Leica TCS SP8 
STED laser confocal microscope is shown in Fig. \ref{fig9}(b). Fig. \ref{fig9}(c) shows 
fluorescence imaging of BEAS-2B of the same sample by laser excitation with wavelength of 
405 nm. Fig. \ref{fig9}(d) is the microscopic image of BEAS-2B with the combination of 
Fig. \ref{fig9}(b) and (c). We can obtain that the BEAS-2B could still have good cell 
activity and fluorescence property with an appropriate CDs concentration. Meanwhile, 
it had been shown that positively charged CQDs are more cytotoxic and have lower photoluminescence 
(PL) compared to negative CQDs \cite{Yan18}. After CDs passivation, its surface positive 
charge is reduced and its toxicity is reduced too.

\section{Conclusions}
In this study, the fluorescent CDs were extracted from cigarette smoke
and passivated by sodium borohydride to enhance the fluorescence of the CDs.
The CDs were applied to improve the imaging quality of biological cells.
The average size of the core aromatic structure region of $sp^2$/$sp^3$ carbon 
is about 3 nm and its edge contains abundant hydroxyl, carbonyl, carboxyl and 
other oxygen-containing functional groups. Under the excitation wavelength of 
360 nm, the fluorescence peak wavelength is 456 nm and the fluorescence quantum 
yield is 6.13\%. The fluorescence intensity of R-CDs obtained by reducing CDs 
with sodium borohydride was increased by about 7 times and the fluorescence 
yield was increased to 8.86\%. Meanwhile, the fluorescence peak has red-shift 
to 463 nm. When R-CDs drops were added to the surface of onion epidermal cells 
and absorbed naturally, it could be clearly observed that the cell wall and 
nucleus emit bright fluorescence from R-CDs. CDs also show low toxicity to 
BEAS-2B with good biological activity. These results indicate that CDs and R-CDs 
have good biocompatibility and fluorescence properties which can improve the 
imaging quality of biological cells.

\begin{backmatter}
\bmsection{Funding}
National Natural Science Foundation of China (12004331, 12064049, 62175209); Yunnan Provincial
Science and Technology Department (202004AP080053); Yunnan Provincial Department of Education (202210691029,
2022J0440).
\bmsection{Disclosures}
The authors declare no conflicts of interest.
\bmsection{Data Availability Statement}
Data underlying the results presented in this paper are not publicly available at this time but may
be obtained from the authors upon reasonable request.
\end{backmatter}


\end{document}